\documentclass{ifacconf}

\usepackage{graphicx}      
\usepackage{natbib}        
\usepackage{xcolor}
\usepackage{comment}
\usepackage{subfigure}
\usepackage{amsmath} 
\usepackage{amssymb}  
\newtheorem{definition}{Definition}

\newtheorem{theorem}{Theorem}
\newtheorem{property}{Property}

\usepackage{lipsum}
\usepackage{xcolor}

\usepackage[most]{tcolorbox}
\newtcolorbox{highlighted}{colback=yellow,coltext=red,breakable}

\def\maxgen#1#2{\left\lceil#1\right\rceil_{#2}}

\def\conv#1#2{ \left( \begin{array}{c} #1 \\ #2 \\
\end{array}\right)} 

\newcommand{\blista}{\renewcommand{\labelenumi}{(\roman{enumi})} 
\begin{enumerate}}
\newcommand{\elista}{\end{enumerate} \renewcommand{\labelenumi}{\arabic{enumi}.}}

\def\setW{{\mathcal{W}}}
\def\setD{{\mathcal{D}}}
\def\setV{{\mathcal{V}}}
\def\setX{{\mathcal{X}}}
\def\setU{{\mathcal{U}}}

\newcommand {\bsis} {\left\{ \begin{array} }
\newcommand {\esis} {\end{array}\right.}

\newcommand{\QED}{$\Box$}

\def\vv#1{{ \rm \bf{#1}}}

\def\Pr{{\mathbb{P}}}
\def\R{ {\rm \,I\!R} }
\def\N{ {\rm \,I\!N} }   

\def\Sum#1#2{\sum\limits_{#1}^{#2}}

\def\bmat#1{\left[\begin{array}{#1}}
\def\emat{\end{array}\right]}

\def\violation#1{\langle #1\rangle_+ }

\begin{document}
\begin{frontmatter}

\title{A probabilistic validation approach for penalty function design in \\ Stochastic Model Predictive Control} 

\thanks[footnoteinfo]{Fabrizio Dabbene aknowledges the Italian Institute of Technology and the Italian Ministero dell'Istruzione, dell'Universit\`a e della Ricerca (PRIN 2017 N. 2017S559BB). Teodoro Alamo acknowledges MEyC Spain (contract DPI2016-76493-C3-1-R).\\ $^1$ corresponding author.}

\author[First]{Martina Mammarella}\footnote{corresponding author.} 
\author[Second]{Teodoro Alamo} 
\author[Third]{Sergio Lucia}
\author[First]{Fabrizio Dabbene}

\address[First]{Institute of Electronics, Computer and Telecommunication Engineering, National Research Council of Italy, Turin, Italy (e-mail: martina.mammarella@ieiit.cnr.it, fabrizio.dabbene@ ieiit.cnr.it).}
\address[Second]{Departamento de Ingenier\'ia de Sistemas y Autom\'atica, Universidad de Sevilla, Escuela Superior de Ingenieros, Camino de los Descubrimientos s/n, 41092 Sevilla, Spain (e-mail: talamo@us.es)}
\address[Third]{Einstein Center Digital Future, Technische Universität Berlin, Germany (e-mail: sergio.lucia@tu-berlin.de)}

\begin{abstract}                
In this paper, we consider a stochastic Model Predictive Control able to account for effects of additive stochastic disturbance with unbounded support, and requiring no restrictive assumption on either independence nor Gaussianity. We revisit the rather classical approach based on penalty functions, with the aim of designing a control scheme that meets some given probabilistic specifications. The main difference with previous approaches is that we do not recur to the notion of probabilistic recursive feasibility, and hence we do not consider separately the unfeasible case. In particular, two probabilistic design problems are envisioned. The first randomization problem aims to design \textit{offline} the constraint set tightening, following an approach inherited from tube-based MPC. For the second probabilistic scheme, a specific probabilistic validation approach is exploited for tuning the penalty parameter, to be selected \textit{offline} among a finite-family of possible values. The simple algorithm here proposed allows designing a \textit{single} controller, always guaranteeing feasibility of the online optimization problem. The proposed method is shown to be more computationally tractable than previous schemes. This is due to the fact that the sample complexity for both probabilistic design problems depends on the prediction horizon in a logarithmic way, unlike scenario-based approaches which exhibit linear dependence. The efficacy of the proposed approach is demonstrated with a numerical example.
\end{abstract}

\begin{keyword}
Predictive control, randomized algorithms, sampling methods, stochastic systems, optimization.
\end{keyword}
\end{frontmatter}
\section{Introduction}\label{sec:intro}
Model predictive control (MPC) is a popular control strategy mainly for its ability to deal with multivariate systems and constraints in a systematic fashion.
However, the presence of uncertainties can significantly degrade closed-loop performance, cause violation of constraints or even lead to instabilities.
These shortcomings have been addressed by many research works, since the first formulation of robust MPC schemes \citep{campo1987} based on worst-case analysis. Indeed, traditional robust MPC schemes minimize the chosen cost function for the worst-case value of the uncertainties, which are assumed to be defined in a compact set, and enforce the constraints for all possible realizations of the uncertainties. On the other hand, because the worst-case value of the uncertainties can have a very small probability of occurrence and any knowledge about their probability distribution is ignored, traditional robust MPC schemes can be very conservative.

An alternative to mitigate such conservativeness is to formulate stochastic MPC problems \citep{mesbah2016stochastic}, which explicitly consider probability distribution functions, including expectations or standard deviations in the cost functions as well as the use of constraints that should be fulfilled in probability, often called \emph{chance constraints}.
Stochastic MPC formulations face two important challenges: (i) the propagation of the stochastic uncertainty through the system dynamics; and (ii) the consideration of chance constraints and recursive feasibility. Unfortunately, exact formulations are intractable even in the linear case. Hence, previous works have been focused on different simplifying assumptions, such as conservative approximation of chance constraints based on the propagation of the variance through linear dynamics (\cite{farina2016-aut}, \cite{hewing2018stochastic}) or via polynomial chaos expansions \citep{mesbah2016stochastic}. Other formulations use the scenario approach (\cite{prandini2012randomized}, \cite{calafiore2013stochastic}, \citep{schildbach2014scenario}, \cite{margellos2014road}), and also an offline scenario setting that leads to a reduced number of samples and lower computational load as shown in \cite{lorenzen2017offline},  \cite{Mammarella:18:Control:Systems:Technology}.
Another problem to deal with in a stochastic MPC setting is related to guaranteeing recursive feasibility. To overcome this difficulty, the approach presented in \cite{lorenzen2017offline} employs constraint tightening to guarantee robust constraint satisfaction for bounded uncertainty whereas \cite{fleming2019-tac} uses first-step chance constraint and applies robust constraints for the rest of the horizon.

The main contribution of this work is the achievement of the desired closed-loop guarantees by means of probabilistic validation techniques \citep{TeBaDa:97}, \citep{Alamo:15}, which are used at two different levels. The first use of probabilistic validation is to compute \textit{offline} a constraint tightening, following the stochastic tube-based MPC approach proposed in \citep{lorenzen2016constraint}, but using probabilistic validation setting instead of the scenario approach. Secondly, to guarantee recursive feasibility, we relax the constraints using a penalty function method \cite{kerrigan2000soft} and, following ideas presented in \cite{karg2019probabilistic}, we perform an offline probabilistic design of the penalty parameter, selected among a finite-family of values, so that the desired probabilistic guarantees of the closed-loop constraint satisfaction are fulfilled.

The proposed approach leads to an MPC formulation that is always feasible and it can obtain a verifiable closed-loop performance. An important merit of our method is that no assumptions on independence or Gaussianity of the stochastic variables are necessary. In addition, the obtained sample complexity does not depend on the design space as in the scenario approach \citep{calafiore2013stochastic}, or on quantities difficult to compute in general such as the Vapnik–Chervonenkis (VC) dimension \citep{lorenzen2017offline}. As a result, the sample complexity depends on the prediction horizon only in a logarithmic way, significantly reducing the number of samples to draw and consequently the computational load.

The remainder of the paper is organized as follows. Section~\ref{sec:prob_setup} describes the mathematical problem setup. Section~\ref{sec:upper_bound_gen} bounds the effect of the disturbances while Section~\ref{sec:constr_tight} describes how to design a proper tightening of the constraints. Section~\ref{sec:penalty} describes the penalty function method used to obtain always a feasible optimization problem and Section~\ref{sec:rho_sample} discusses the choice of the penalty parameter. Finally, the potential of the approach is shown via a numerical example in Section~\ref{sec:results} while main conclusions of the work are presented in Section~\ref{sec:conclusion}.

{\textit{Notation}: The set $\mathbb{N}_{>0}$ denotes the positive integers, the set $\mathbb{N}_{\geq 0} = \left\{0\right\} \cup\mathbb{N}_{>0}$ the non-negative integers, and $\mathbb{N}_a^b$ the integers interval $[a,b]$. Similarly $\mathbb{R}_{>0}$ ($\mathbb{R}_{\geq0}$) for positive real numbers. We use $x_k$ for the (measured) state at time $k$ and $x_{\ell|k}$ for the state predicted $\ell$ steps ahead at time $k$. Positive (semi)definite matrices $A$ are denoted $A\succ 0$ $(A\succeq 0)$ and $\|x\|_A^2\doteq x^TAx$. For vectors, $x\succeq 0$ ($x\preceq 0$) is intended component-wise. Calligraphic upper-case letters, e.g. $\mathcal{A}$, denote sets.  $\Pr_\mathcal{A}$ denotes the probabilistic distribution of a random variable $a\in\mathcal{A}$. Sequence of scalars/vectors are denoted with bold lower-case letters, i.e. $\textbf{v}$. Given a vector $\alpha=[\alpha_1,\ldots,\alpha_{n_\alpha}]^{T}\in\mathbb{R}^{n_\alpha}$, then $\violation{\alpha}$ is a scalar defined as $ \violation{\alpha}\doteq \Sum{i=1}{n_\alpha} \max\{ 0, \alpha_i\}$.}


\section{Problem Setup}
\label{sec:prob_setup}
Let us consider the following linear time-invariant system affected by persistent, additive disturbance $\zeta_k \in \mathbb{R}^{n_x}$ 
\begin{equation}
    x_{k+1}=Ax_k+Bu_k+\zeta_k,\; \forall k\in\mathbb{N}_{\geq0}
    \label{eq:sys}
\end{equation}

where $x_k\in \mathbb{R}^{n_x}$ is the state variable at time $k$, $u_k \in \mathbb{R}^{n_u}$ is the control input, and $A$ and $B$ are matrices of appropriate dimensions. No assumption on neither independence nor Gaussianity are made on the stochastic disturbance $\zeta_k$. Moreover, both state and input are constrained in compact sets $\setX$ and $\setU$, respectively, and the corresponding constraints can be defined in a compact form as
\begin{equation}
    Cx_k+Du_k\preceq h.
    \label{eq:constr}
\end{equation}
%
%
The control objective is to design a stabilizing receding horizon control, which guarantees constraint satisfaction in a probabilistic setting. We will consider the following quadratic stage cost
\begin{equation}
    L(x_k,u_k) \doteq \| x_k\|_Q^2+\|u_k\|_R^2,
    \label{eq:cost_L}
\end{equation}
where $Q\in\mathbb{R}^{n_x\times n_x}$, $Q\succeq 0$, $R\in\mathbb{R}^{n_u\times n_u}$, $R\succ 0$.
To solve the control problem, a stochastic MPC algorithm is considered where, as typical of predictive schemes, the optimal control problem is solved repeatedly over a finite horizon $N$, but only the first control action of the optimal sequence is implemented (see \cite{mayne2000constrained} for a meticulous review on MPC). The proposed controller is designed by means of a two-step procedure:
\blista
\item Using a sampling method, we first bound the effect of disturbances in a probabilistic manner. This allows us to formulate a nominal model predictive controller that addresses the chance constraints issue (Section \ref{sec:upper_bound_gen} and Section \ref{sec:constr_tight}). 
\item In order to avoid infeasibility of the proposed MPC approach, we rewrite the controller using a penalty cost scheme. In particular, the penalty factor is adjusted using sampling in such a way that the resulting controller meets online the probabilistic specifications on the constraints satisfaction (Section \ref{sec:penalty} and Section \ref{sec:rho_sample}).  
\elista

\section{Probabilistic upper bounds of the effect of disturbances}
\label{sec:upper_bound_gen}
As it is common in robust and stochastic MPC, let us consider the state of the system $x_{\ell|k}$, predicted $\ell$ steps ahead from time $k$, split into a deterministic, nominal part $z_{\ell|k}$ and an error part $e_{\ell|k}$ as
\begin{equation}
    x_{\ell|k}=z_{\ell|k}+e_{\ell|k}.
    \label{eq:state_split}
\end{equation}
Then, a parametrized feedback policy of the form
\begin{equation}
    u_{\ell|k}=v_{\ell|k}+Kx_{\ell|k},\; \forall \ell\in\mathbb{N}_{0}^{N-1},
    \label{eq:input_law}
\end{equation}
is considered where the feedback gain matrix $K$ is quadratically stabilizing for the system
\eqref{eq:sys}.

Hence, considering the feedback policy \eqref{eq:input_law}, the system dynamics in \eqref{eq:sys} along the prediction horizon $N$ can be rewritten in terms of nominal and error dynamics as
\begin{subequations}
    \begin{align}
        z_{\ell+1|k}&=A_Kz_{\ell|k}+Bv_{\ell|k},\; z_{0|k}=x_k,
        \label{eq:nom_state}
        \end{align}
        \begin{align}
            e_{\ell+1|k}&=A_Ke_{\ell|k}+\zeta_{\ell|k},\; e_{0|k}=0,
        \label{eq:err_state}
    \end{align}
\end{subequations}
where $\zeta_{\ell|k}\doteq \zeta_{\ell+k}$ and $A_K\doteq A+BK$. 
%
Now, considering the state decomposition in \eqref{eq:state_split} and the feedback policy \eqref{eq:input_law}, the constraint \eqref{eq:constr} can be rewritten as 
\begin{equation}
    C_Kz_{\ell|k}+Dv_{\ell|k}+C_Ke_{\ell|k} \preceq h, \; \forall \ell\in \N_{\ell=0}^{N-1}
    \label{eq:constr_1}
\end{equation}
where $C_K\doteq (C+DK)$.
In absence of disturbances, $e_{\ell|k} =0$, for all $\ell \in \N_{\ell=0}^{N-1}$ and, consequently, the constraints given in (\ref{eq:constr_1}) would be equivalent to 
\begin{equation}\label{ineq:nominal:constraints}
    C_Kz_{\ell|k}+Dv_{\ell|k}\preceq h, \; \forall \ell\in \N_0^{N-1}. 
\end{equation}   
On the other hand, in the presence of random disturbances, one has to deal with the uncertain random vectors $C_K e_{\ell|k}$, with $\ell\in \N_{\ell=0}^{N-1}$ that appear in (\ref{eq:constr_1}). One possibility is to probabilistically upper bound those terms. This is precisely the objective of the remaining of this section. 

\subsection{Preliminaries: Probabilistic upper bound of a random variable}

We first present a generalization of the notion of the maximum of a collection of scalars, borrowed from the field of order statistics \citep{Ahsanullah:13,Arnold:92}. This will allow us to reduce the conservativeness that follows from the use of the standard notion of max function. See also Section 3 of \cite{Alamo:18}.\\

\begin{definition}[Ordered Sequence]\label{def:def1}
Given a sequence of $S$ scalars $$ \vv{v} = \{ v^{(1)}, v^{(2)}, \ldots , v^{(S)}\} = \{v^{(i)}\}_{i=1}^S,$$ we denote with $\vv{v}_+ = \{v_+^{(i)}\}_{i=1}^S $ the ordered sequence obtained by rearranging the elements of $\vv{v}$ in a non-increasing order. That is 
\[
v_+^{(1)} \geq v_+^{(2)} \geq \ldots \geq v_+^{(S-1)} \geq v_+^{(S)}.
\]
\hspace{0.5cm}
\end{definition}

\begin{definition}[Generalized max function $\maxgen{\cdot}{r}$]
Given a sequence $\vv{v} = \{v^{(i)}\}_{i=1}^S$ of $S$ scalars and the integer $r\in[1,S]$, we define the generalized max function $\maxgen{\vv{v}}{r}$ as  
$$ \maxgen{\vv{v}}{r} \doteq  v_+^{(r)}.$$ 
where $\{v_+^{(i)}\}_{i=1}^S$ is given in Definition \ref{def:def1}. 
\end{definition}

Clearly, applying Definition \ref{def:def1},
we have 
$$ \maxgen{\vv{v}}{1} = v_+^{(1)} = \max\limits_{1\leq i\leq S}\,v^{(i)}, \; \maxgen{\vv{v}}{S} = v_+^{(S)} =\min\limits_{1\leq i\leq S}\,v^{(i)}.$$
Furthermore,  $\maxgen{\vv{v}}{2}$ denotes the second largest value in $\vv{v}$, $\maxgen{\vv{v}}{3}$ the third largest one, etc. We notice that the notation $\maxgen{\vv{v}}{r}$ does not need to make explicit $S$, the number of components of $\vv{v}$. 
The following property, which has been already proved in \cite[Property 3]{Alamo:18}, states that the generalized notion of max function can be used to provide a probabilistic upper bound of a given random variable.\\

\begin{property} \label{prop:max:gen} Consider a random scalar variable $v\in \setV$ with probabilistic distribution $\Pr_\setV$. Suppose that  $\vv{v}=\{v^{(i)}\}_{i=1}^S$ is a sequence of $S$ independent identically distributed (i.i.d.) scalars that have been drawn according to $\Pr_\setV$. Then, with probability no smaller than $1-\delta$, 
$$ \Pr_\setV \{ v > \maxgen{\vv{v}}{r}\} \leq \epsilon,  $$
provided that $r\in \N_{1}^{S}$ and 
\begin{equation}\label{ineq:binomial:original}
\Sum{m=0}{r-1}\conv{S}{m}\epsilon^m(1-\epsilon)^{S-m}\leq \delta.
\end{equation}
Moreover, (\ref{ineq:binomial:original}) is satisfied if 
$$ S \geq \frac{1}{\epsilon}\left(  r-1 + \ln\frac{1}{\delta} + \sqrt{2(r-1)\ln \frac{1}{\delta}}\right).
$$
\end{property}
Property \ref{prop:max:gen} has been already used in the context of probabilistic scaling and validation (see \cite{alamo2019safe} and \cite{karg2019probabilistic}). See also \cite{TeBaDa:97} for the particularization of the result to the case $r=1$ and a single constraint. In the following section, we will generalize this result in such a way that it will allow us to address the probabilistic tightening of the control constraints in order to cope with the uncertain disturbances.

\subsection{Sample-based probabilistic upper bounds of the effect of disturbances}
\label{sec:upper_bound}
From the uncertain dynamics given in (\ref{eq:err_state}), we have that the sequence 
$\{e_{\ell|k}\}_{\ell=0}^{N-1}$ is completely determined by the sequence $$ \boldsymbol{\zeta} = \{\zeta_{0|k},\zeta_{1|k},\ldots, \zeta_{N-1|k}\} = \{\zeta_{k},\zeta_{k+1},\ldots, \zeta_{k+N-1}\}.$$
We assume that $\boldsymbol{\zeta}$ is a stationary random vector with probability distribution $\Pr_\setD$ in $\R^{n_x\times N}$.
Since we assume that the probability distribution of $\boldsymbol{\zeta}\in\setD$
is independent of sample time $k$ due to the stationary nature of random sequence $\boldsymbol{\zeta}$, we have that the probability distribution of $e_{\ell|k}$ is equal to the probability distribution of $e_{\ell|0}$. In order to make explicit the dependence of $e_{\ell|k}$ on the uncertain disturbances, we denote by  $\{e_{\ell|k}(\boldsymbol{\zeta})\}_{\ell=0}^{N-1}$ the sequence of error dynamics (\ref{eq:err_state}) corresponding to the sequence $\boldsymbol{\zeta}$. 

Now, we generalize the results of Property \ref{prop:max:gen} to obtain sample-based probabilistic upper bounds for 
$$ C_{K,j}e_{\ell|k}(\boldsymbol{\zeta}), \; j\in\mathbb{N}_1^{n_h}, \; \ell \in\mathbb{N}_0^{N-1},$$
where $C_{K,j}$ denotes the $j$-th row of matrix $C_{K}$.\\

\begin{theorem}
\label{th:theo1}
Given a discarding parameter $r_q$, and the probabilistic levels $\epsilon_q\in(0,1)$ and $\delta_q \in(0,1)$, suppose that $S_q$ i.i.d. samples  $\{\boldsymbol{\zeta}^{(1)}, \ldots, \boldsymbol{\zeta}^{(S_q)} \}$ are drawn according to $\Pr_\setD$. Let us assume also that the vectors 
$$\{ q_0,q_1, \ldots, q_{N-1}\}\in \R^{n_h\times N}$$  are computed using the following expression
\begin{equation}
    q_{\ell,j} = \maxgen{\{ C_{K,j}e_{\ell|0}(\boldsymbol{\zeta}^{(i)})\}_{i=1}^{S_q}}{r_q}, \forall \ell\in \N_{0}^{N-1}, \,\forall j \in \N_1^{n_h},
    \label{eq:ql}
\end{equation}
where $q_{\ell,j}$ is the $j$-th component of $q_\ell\in \R^{n_h}$. Then, with probability no smaller than $1-\delta_q$, we have
$$\Pr_\setD \{C_{K,j} e_{\ell|k}(\boldsymbol{\zeta}) > q_{\ell,j}\} \leq \epsilon_q, \; \forall \ell\in \N_0^{N-1}, \; \forall j \in \N_1^{n_h},$$

provided that $r\in\mathbb{N}_1^{S_q}$ and
\begin{equation}
\label{eq:binomial_ineq}
  \sum_{m=0}^{r_q-1} \conv{S_q}{m}\epsilon_q^m (1-\epsilon_q )^{Sq-m} \leq \frac{\delta_q}{n_h N}.   
 \end{equation}

In addition, \eqref{eq:binomial_ineq} is satisfied if
\begin{equation}
S_q \geq \frac{1}{\epsilon_q}\left(  r_q-1 + \ln\frac{n_h N}{\delta_q} + \sqrt{2(r_q-1)\ln \frac{n_h N}{\delta_q}}\right).
\label{eq:ns_for_qk}
\end{equation}
\end{theorem}

Proof to \textit{Theorem \ref{th:theo1}} can be found in Appendix \ref{app:app1}.

\section{Sample-based constraint tightening}
\label{sec:constr_tight}
The objective of this section is  to present a model predictive controller with a constraint tightening based on the probabilistic upper bounds 
$\{q_\ell\}_{\ell=0}^{N-1}$ that can be obtained from Theorem \ref{th:theo1}.

For a classical MPC scheme with semi-feedback structure $u_{\ell|k}=Kx_{\ell|k}+v_{\ell|k}$,  
the finite horizon cost $J(\textbf{x}_{k},\textbf{v}_{k})$ to be minimized at time $k$ is defined as 
\begin{equation}
    J(\textbf{x}_{k},\textbf{v}_{k})=\sum_{\ell=0}^{N-1}(\|x_{\ell|k}\|_Q^2+\|Kx_{\ell|k}+v_{\ell|k}\|_R^2)+\|x_{N|k}\|_{\tilde{P}}^2,
    \label{eq:J_classic}
\end{equation}
where $\textbf{x}_k=[x_{0|k},\ldots,x_{N|k}]^{\text{T}}$, $\textbf{v}_k=[v_{0|k},\ldots, v_{N-1|k}]^{\text{T}}$, and $\tilde P$ is the solution of the discrete-time Riccati equation
\begin{equation}
    Q+K^{\text{T}}RK+A_K^{\text{T}}\tilde PA_K = \tilde P.\\
    \label{eq:riccati}
\end{equation}
Then, a nominal finite horizon optimization problem $P_{nom}(x_k)$ can be defined as
\begin{subequations}\label{eq:Pt_x}
\begin{align}
\min\limits_{\textbf{x}_k,\textbf{v}_k}\,\,\,\,& J(\textbf{x}_k,\textbf{v}_k)\\
 s.t.\,\,\,\,\, & x_{0|k} = x_k, \\
 \,\,\,\,& x_{\ell+1|k} = A_Kx_{\ell|k}+Bv_{\ell|k}\; \forall \ell\in\mathbb{N}_0^{N-2}, \\
\,\,\,\, & x_{N-1|k} = A_K x_{N-1|k} +Bv_{N-1|k}, \label{equ:equilibrio} \\
\,\,\,\, & C_K x_{\ell|k} +Dv_{\ell|k} \preceq h,\; \forall \ell\in \mathbb{N}_0^{N-1}.\label{eq:constraints_Px}
\end{align}
\label{eq:Px_x}
\end{subequations}
We notice that constraint (\ref{equ:equilibrio}) is a terminal constraint that forces $x_{N-1}$ to be an equilibrium point for the nominal system. 
In this work, we inherit the typical approach exploited in tube-based MPC schemes (see e.g. \cite{raw}), where the control objective becomes controlling the nominal dynamics $z_{\ell|k}$ in \eqref{eq:nom_state} by solving the following optimization problem subject to a \textit{tightened} version of the nominal constraints given in \eqref{ineq:nominal:constraints}.\\

\begin{definition}\label{def:P:q} (\textit{Finite Horizon Optimization Problem with Tightened Constraints}) Given an initial condition $z_{0|k}\in\mathbb{R}^{n_x}$, with $z_{0|k}=x_k$, we formulate the optimization problem $P_q(x_k)$ with tightened constraints as
\begin{subequations}\label{eq:Pt_x}
\begin{align}
\min\limits_{\textbf{z}_k,\textbf{v}_k}\,\,\,\,& J(\textbf{z}_k,\textbf{v}_k)\\
 s.t.\,\,\,\, & z_{0|k} = x_k,\label{eq:15b}\\
\,\,\,\, & z_{\ell+1|k} = A_Kz_{\ell|k}+Bv_{\ell|k} ,\; \forall \ell\in\mathbb{N}_0^{N-2}, \label{eq:15c} \\
\,\,\,\, & z_{N-1|k} = A_K z_{N-1|k} +Bv_{N-1|k},\label{eq:15d} \\
\,\,\,\, & C_K z_{\ell|k} +Dv_{\ell|k} \preceq h-q_{\ell},\; \forall \ell\in \mathbb{N}_0^{N-1},\label{eq:constraints_Px}
\end{align}
\label{eq:Px}
\end{subequations}
We denote with $\setX_N$ the set of initial conditions $x_k$ for which problem $P_q(x_k)$ is feasible. For every $x_k\in \setX_N$ we denote the minimizer of $P_q(x_k)$ by  $(\textbf{z}_k^*,\textbf{v}_k^*)=(z_{0|k}^*,\ldots,z_{N|k}^*,v_{0|k}^*,\ldots,v_{N-1|k}^*)$.
\end{definition}

It is important to highlight that, as in \eqref{eq:J_classic}, a weighted terminal cost is included to ensure that the optimal cost $J(\textbf{z}_k,\textbf{v}_k)$ is a Lyapunov function for the system, guaranteeing stability. By properly weighting the terminal cost, the domain of attraction of this controller can be enlarged. See e.g. \cite{limon2006stability} and references therein. 

From (\ref{eq:constr_1}) we have that the control constraints 
$$ Cx_{\ell|k} + Du_{\ell|k} \preceq h, \forall \ell\in \N_0^{N-1},$$ can be rewritten as 
\begin{equation}\label{ineq:constraint:with:e}
    C_K z_{\ell|k}+D u_{\ell|k}\preceq h - C_K e_{\ell|k}. 
\end{equation} 
Thus, the tightened constraints (\ref{eq:constraints_Px}) guarantee the satisfaction of (\ref{ineq:constraint:with:e}) provided that $C_k e_{\ell|k} \preceq q_\ell$. We conclude that 
$$ \bsis{rcl} C_K z_{\ell|k} + D v_{\ell|k} &\preceq& h-q_\ell \\
C_k e_{\ell|k} &\preceq& q_\ell \esis \Rightarrow Cx_{\ell|k}+Du_{\ell|k} \preceq h.$$ 
From here we infer that, for every feasible solution $(\textbf{z}_k,\textbf{v}_k)$ of $P_q(x_k)$,
$$ \Pr_\setD\{ Cx_{\ell|k}+Du_{\ell|k} \preceq h \} \geq \Pr_\setD\{ C_K e_{\ell|k} \preceq q_\ell\}.$$
Denoting $C_j$, $D_j$, and $C_{K,j}$ the $j$-th rows of matrices $C$, $D$ and $C_K$ respectively, and $h_j$, $q_{\ell,j}$ the $j$-th components of $h$ and $q_\ell$ we also obtain
$$ \Pr_\setD\{ C_jx_{\ell|k}+D_ju_{\ell|k} \leq h_j \} \geq \Pr_\setD\{ C_{K,j} e_{\ell|k} \leq q_{\ell,j}\}.$$
Given $\delta_q\in(0,1)$ and $\epsilon_q\in(0,1)$, Theorem \ref{th:theo1} provides a way to obtain $\{q_\ell\}_{\ell=0}^{N-1}$ such that, with probability no smaller than $1-\delta_q$,
$$\Pr_\setD \{C_{K,j} e_{\ell|k}(\boldsymbol{\zeta}) > q_{\ell,j}\} \leq \epsilon_q, \; \forall \ell\in \N_0^{N-1}, \; \forall j \in \N_1^{n_h}.$$
Thus, we conclude that, given $\delta_q\in(0,1)$ and $\epsilon_q\in(0,1)$, if $\{q_\ell\}_{\ell=0}^{N-1}$ are obtained according to Theorem \ref{th:theo1} then, with probability no smaller than $1-\delta_q$, 
\[
\Pr_\setD\{ C_j x_{\ell|k}+D_ju_{\ell|k} \leq h_j \} \geq 1-\epsilon_q, \; \forall \ell\in \N_0^{N-1}, \; \forall j \in \N_1^{n_h}.
\]
We notice that the previous probabilistic bound does not refer to the closed-loop behaviour, but to the prediction scheme of the stochastic MPC formulation. In the following sections, we present how to design a soft constrained version of the controller proposed in (\ref{eq:Pt_x}) in order to obtain closed-loop probabilistic guarantees. 

\section{Soft-constrained  controller}
\label{sec:penalty}

The feasibility region $\setX_N$ of problem $P_q(\cdot)$ is often a bounded region around the origin. Moreover, if the probability distribution of the disturbances has not a finite support, then the on-line recursive feasibility of $P_q(\cdot)$ can only be guaranteed in a probabilistic way (see, for example, \cite{fleming2019-tac}). In order to circumvent this problem, we propose a soft constrained formulation of the optimization problem that defines the stochastic MPC controller. The proposed scheme relies on the notion of penalty function (\cite{kerrigan2000soft}). Given a penalty factor $\rho>0$, and an initial condition $x_k$, we define the new optimization problem $P_\rho(x_k)$ as
\begin{subequations}
\begin{align}
\min\limits_{\textbf{z}_k,\textbf{v}_k}\,\,\,\,& J(\textbf{z}_k,\textbf{v}_k) + \rho \Sum{\ell=0}{N-1} \violation{C_K z_{\ell|k} +Dv_{\ell|k} - h + q_\ell}\\
 s.t.\,\,\,\, & \eqref{eq:15b},\;\eqref{eq:15c},\;\eqref{eq:15d},\\
\,\,\,\, & u_{0|k} = v_{0|k}+Kx_{0|k}\in \setU.\label{eq:Pt_x_2}
\end{align}
\end{subequations}

To guarantee that the controller provides admissible control action, we have explicitly added a first step constraint on the input $u_{0|k}$ in \eqref{eq:Pt_x_2}, defining the feasible first inputs of the finite horizon program. Under very general assumptions (controllability and $N\geq n_x$), problem $P_\rho(x_k)$ is always feasible.   
%
%
Moreover, $P_\rho(x_k)$ can be cast into the following equivalent optimization problem using a slack variable $\eta$
\begin{subequations}
    \begin{align}
    \min\limits_{\textbf{z}_k,\textbf{v}_k,\eta}\,\,\,\,& J(\textbf{z}_k,\textbf{v}_k) + \rho \|\eta\|_1 \label{eq:J_final} \\
 s.t.\,\,\,\,& z_{0|k} = x_k,\\
 \,\,\,\, & z_{\ell+1|k} = A_Kz_{\ell|k}+v_{\ell|k} ,\; \forall \ell\in\mathbb{N}_0^{N-2},\\
 \,\,\,\, & z_{N-1|k} = Az_{N-1|k}+Bv_{N-1|k},\\
 \,\,\,\,& u_{0|k} = v_{0|k}+Kx_{0|k}\in \setU,\\
 \,\,\,\,& C_K z_{\ell|k} +Dv_{\ell|k} - h + q_\ell \preceq \eta,\;\forall \ell\in\mathbb{N}_0^{N-1}\\
 \,\,\,\,& \eta \succeq 0.
     \end{align}
    \label{eq:p_tilde_3}
\end{subequations}

Since the components of $\eta$ are restricted to be non-negative, $\|\eta\|_1$ is equal to the sum of the components of~$\eta$. This implies that optimization problem \eqref{eq:p_tilde_3} amounts to the minimization of a semi-definite quadratic function subject to a set of linear equalities and inequalities. Thus, \eqref{eq:p_tilde_3} is a semi-definite quadratic optimization problem, for which there exits many reliable solvers suited for fast-embedded implementations, e.g. \textsc{OSQP} and \textsc{CVXGEN}. See \cite{stellato2018osqp}, \cite{mattingley2012cvxgen} and references therein.


 \section{Sampled-Based Design of the penalty factor $\rho$}
\label{sec:rho_sample}

The objective of this section is to determine, by means of a sampled-based scheme, a value of the penalty factor $\rho$ able to provide some probabilistic guarantees with respect to the {\textit closed-loop} satisfaction of \eqref{eq:constr} along a given simulation horizon $M$
$$ Cx_k+Du_k \preceq h, \; k=0,\ldots,M.$$
 
Givens $\rho\in(0,\infty)$ and the state vector $x_k$, the control action $u_k$ corresponding to the control scheme presented in Section \ref{sec:penalty} is $u_k=v_{0|k}^*+Kx_k$, where $v_{0|k}^*$ is the first element of the optimal control sequence $\textbf{v}_k^*$, solution of \eqref{eq:p_tilde_3}. The resulting controller is defined as $u_k=\kappa(x_k,\rho)$, where we make explicit the penalty factor $\rho$. Hence, the closed-loop system \eqref{eq:sys} becomes  
  $$ x_{k+1} = Ax_k+B\kappa(x_k,\rho)+\zeta_k, \forall k\in\N_{\geq 0}.$$

In order to characterize the closed-loop behaviour of the system, we consider a simulation horizon $M$ considerably larger than the prediction horizon $N$ used in the definition of the controller $\kappa(\cdot,\rho)$. In particular, we assume that the horizon $M$ is large enough to guarantee that the state $x_M$ reaches a safe region around the reference steady state. Thus, the closed-loop trajectory $\{ x_k,u_k\}_{k=0}^M$ is determined by 
 \blista
 \item the penalty factor $\rho$;
 \item the initial condition at $k=0$, i.e. $x_0$;
 \item the uncertain realization of the disturbances $$\textbf{d}_M=\{\zeta_0,\ldots, \zeta_{M-1}\}.$$
 \elista

Now, let us consider a  probability distribution in $\setX_N$, i.e. the feasibility region of optimization problem (\ref{eq:Pt_x}). Moreover, given the simulation horizon $M$, we define $\setD_M$ as the set of the possible values for $\textbf{d}_M$ and a probability distribution in it too. In order to simplify the notation, we define 
 $\setW$ as the set of possible values for 
 $$ w=\{x_0,\textbf{d}_M\}=\{x_0,\zeta_0, \ldots, \zeta_{M-1}\} \in \setX_N\times \setD_M=\setW.$$ 
We assume that we are able to draw i.i.d. samples from~$\setW$. Then, the closed-loop trajectory corresponding to controller $\kappa(\cdot,\rho)$ and uncertain realization $w$ is denoted as
 $$\{ x_k(w,\rho),u_k(w,\rho)\}_{k=0}^{M}.$$
 
Given the uncertain realization $w$ and $\rho$, we can determine if the corresponding closed-loop trajectory satisfies the control constraints by means of the computation of the following performance index
 $$ g(w,\rho) \doteq \Sum{k=0}{M} \violation{ Cx_k(w,\rho)+Du_k(w,\rho)-h}.$$


Hence, we have
$$ g(w,\rho)=0 \Leftrightarrow Cx_k(w,\rho)+Du_k(w,\rho)\preceq h, \; \forall k\in \N_{0}^M,$$
and we can conclude that $g(w,\rho)$ serves as an index to evaluate to which extent the control constraints have been satisfied along the closed-loop trajectory.

Now, we consider that $\rho$ is allowed to take values from a set of finite cardinality $\boldsymbol{\Theta}_\rho=\{\rho_1,\rho_2,\ldots,\rho_{n_C}\}$, where $n_C$ is the cardinality of $\boldsymbol{\Theta}_\rho$. The next theorem presents a sample-based scheme that, for any $\rho\in \boldsymbol{\Theta}_\rho$, provides a probabilistic upper bound on $g(w,\rho)$.\\

\begin{theorem}\label{theo:probawilistic:rho}
Let us suppose to draw $S_\rho$ i.i.d. samples of $w^{(i)}$ from $\setW=\setX_N\times\setD_M$, i.e.  
$$ \{w^{(1)}, \ldots, w^{(S_\rho)}\},$$ 
and that $r_\rho$ is a given discarding parameter. Moreover, given any $\rho\in \boldsymbol{\Theta}_\rho$, we introduce the following notation 
\begin{equation}\label{equ:gamma:rho:def}
    \gamma(\rho)  =  \maxgen{\{ g(w^{(i)}, \rho)\}_{i=1}^{S_\rho}}{r_\rho}.
\end{equation} 
Then, with probability no smaller than $1-\delta_\rho$, we have 
\begin{equation}\label{ineq:prob:rho} \Pr_\setW\{g(w,\rho) > \gamma(\rho)\}\leq \epsilon_\rho, \; \forall \rho\in \boldsymbol{\Theta}_\rho,
\end{equation}
provided that $r_\rho\in\N_{1}^{S_\rho}$ and
\begin{equation}
\label{eq:binomial_ineq:rho}
  \sum_{m=0}^{r_\rho-1} \conv{S_\rho}{m}\epsilon_\rho^m (1-\epsilon_\rho )^{S_\rho-m} \leq \frac{\delta_\rho}{n_C}.
 \end{equation}
In addition, \eqref{eq:binomial_ineq:rho} is satisfied if
\begin{equation}
S_\rho \geq \frac{1}{\epsilon_\rho}\left(  r_\rho-1 + \ln\frac{n_C}{\delta_\rho} + \sqrt{2(r_\rho-1)\ln \frac{n_C}{\delta_\rho}}\right).
\label{eq:S:for:rho}
\end{equation}
\end{theorem}

Proof to \textit{Theorem \ref{theo:probawilistic:rho}} follows similar developments to that of \textit{Theorem \ref{th:theo1}} and is omitted for brevity. 

We notice that the probabilistic guarantees given in (\ref{ineq:prob:rho}) are valid for every value of $\rho$ in $\boldsymbol{\Theta}_\rho$. The particular choice for the controller implementation depends on the specific control application. For example, one could choose the value of $\rho$ that minimizes $\gamma(\rho)$ in $\boldsymbol{\Theta}_\rho$. Another possibility is to choose the smallest $\rho$ satisfying a pre-specified constraint on $\gamma(\rho)$. Next section illustrates how to choose~$\rho$ by means of a numerical example.
 

\section{Numerical Example}
\label{sec:results}
In this section, the performances of the proposed stochastic MPC scheme are demonstrated by means of a numerical example previously presented in \cite{lorenzen2016constraint} \footnote{Simulations were run in Matlab R2018a on an Intel Core i7-7500U CPU $@$ 2.9GHz.}. The linear system considered is of the form \eqref{eq:sys} with
\begin{equation*}
    A=\begin{bmatrix}
    1\,\,\,&\,\,\,0.0075\\
    -0.143\,\,\,&\,\,\,0.996
    \end{bmatrix},\;
    B=\begin{bmatrix}
    4.798\\
    0.115
    \end{bmatrix}.
\end{equation*}
For simplicity and coherence with previous works, the disturbance distribution is assumed to be a Gaussian with covariance matrix $\Sigma=0.04^2I_2$ truncated at $\|\zeta\|^2\leq0.02$. Moreover, the SMPC weight matrices are set as $Q=\text{diag}[1 \; 10]$ and $R=1$ whereas the prediction horizon is $N=8$. Consequently, the resulting stabilizing feedback matrix $K=[-0.2858 \; 0.4910]$, solution of the unconstrained discrete LQR problem, is used. The system is subject to hard constraints on the input $|u_k|\leq 0.2$ and chance constraints on the state components $|x_k(1)|\leq 2$ and $|x_k(2)|\leq 3$ (this leads to $n_h=6$ linear constraints), which should be satisfied with probability of at least $95\%$ ($\epsilon_q=0.05$). For the computation of the minimum number of disturbance samples $S_q$  to draw in \eqref{eq:ns_for_qk}, $\delta_q$ has been set equal to $10^{-6}$.  The discarding parameter $r_q$ was selected such that $r_q/S_q \sim \epsilon_q /2$, i.e. $r_q=60$ and $S_q=2448$. 

The starting point consisted in evaluating \textit{offline} the optimal probabilistic constraint tightening $q_{\ell}$. To this end, we first drew $S_q$ i.i.d. samples $\boldsymbol{\zeta}^{(i)}$, $i=1,\ldots,S_q$, from $\setD$  and, for each $i$-th sample, we propagated the error dynamics in \eqref{eq:err_state} to obtain $C_Ke_{\ell|k}$ for each $\ell=\mathbb{N}_0^{N-1}$. Subsequently, we evaluated $q_{\ell,j}$ for $j=\mathbb{N}_1^{n_h}$ as described in Section \ref{sec:upper_bound} obtaining $q_{\ell}~=~[0.0129\;0.0134\;0.0066\;0.0069\;0.0026\;0.0026]^{\text{T}}$. Once assessed the tightening parameter $q_{\ell}$, we moved on to the second probabilistic design problem involving the penalty term $\rho$. First, we set $n_C=100$ finite-family controllers $\kappa(\cdot,\rho_l)$, where the penalty factors $\{ \rho_l\}_{l=1}^{n_C}$ have been obtained taking $n_C$ points in the interval $[1,10^6]$, chosen to be equidistant in a logarithmic scale. That is, 
\begin{equation*}
        \rho_l = \rho_{min}\cdot \exp{\left(\frac{l-1}{n_C-1}\ln{\left(\frac{\rho_{min}}{\rho_{max}}\right)}\right)},\,\,\; l=[1,n_C],
\end{equation*}
with $\rho_{min}=1$ and $\rho_{max}=10^6$.
The probabilistic levels $\epsilon_\rho$ and $\delta_\rho$  have been set equal to $0.05$ and $10^{-6}$, respectively, and with $r_\rho=60$, the sample complexity for the second randomization problem was set to $S_\rho=2614$, according to \eqref{eq:S:for:rho}. Last, the system dynamics was simulated considering a sample time of $\Delta t=1/50$ s and $k=[0,20]$, i.e. simulation horizon $M$ of 20 time steps.
Then, for each controller $\kappa(x_k,\rho_l)$, we evaluated the proposed controller performance simulating the system dynamics \eqref{eq:sys} for $S_\rho$ random scenarios $w$, containing random feasible initial conditions $x_0$ from the feasibly region $\mathcal{X}_N$ (see Definition \ref{def:P:q}) and random disturbance trajectories $\textbf{d}_M$.


%
\begin{figure}[h!]
\centering
\includegraphics[trim=2cm 0.0cm 2cm 0.0cm, clip=true,width=1\columnwidth]{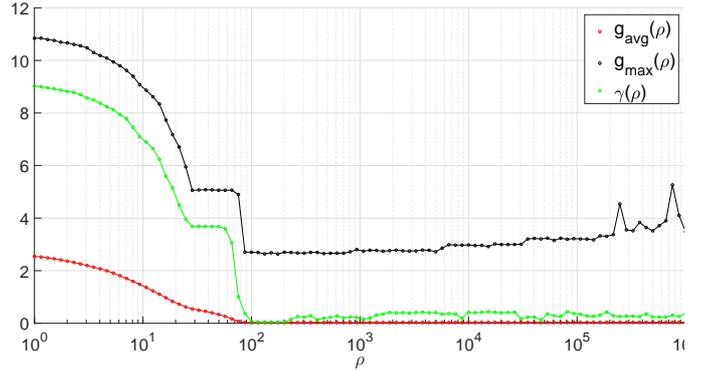}
\caption{Performance index maximum ($g_{max}(\rho)$), average ($g_{avg}(\rho)$) and probabilistic upper bound ($\gamma(\rho)$) over the simulation horizon $M$ for $\rho=[1,10^6]$ and $S_{\rho}=2614$.}
\label{fig:avg_max_22102019}
\end{figure}

The next phase consisted in evaluating, for each of the $S_\rho$ simulations, the corresponding performance index $g(w,\rho)$ over the simulation horizon $M$. Fig. \ref{fig:avg_max_22102019} shows the behavior of three performance indices with respect to the penalty factor $\rho$: i) $g_{avg}(\rho)$ is the average value obtained for $g(w,\rho)$ over the $S_q$ simulations,  ii) $g_{max}(\rho)$ is the largest value and iii) $\gamma(\rho)$ is the generalized maximum of the values of $g(w,\rho)$ using discarding parameter $r_\rho$, see \eqref{equ:gamma:rho:def}. We can observe some similarities among $g_{avg}(\rho)$, $g_{max}(\rho)$ and $\gamma(\rho)$ trends for $\rho$ up to $100$. Indeed, all three curves present a discontinuous decreasing behavior, due to two inflection points for $\rho=30$ and $\rho=80$, before settling around a constant value for higher penalty factors. We also notice that $g_{max}(\rho)$ results more affected by the random nature of the different simulations when compared to $g_{avg}(\rho)$ and $\gamma(\rho)$, which provide more coherent and reliable information about the impact of $\rho$ on constraints violation. Indeed, Fig. \ref{fig:avg_max_22102019} seems to point out that for $\rho\geq 100$ the average violation entity $g_{avg}(\rho)$ settles around $0.035$. On the other hand, $\gamma(\rho)$ presents a local minimum close to a value of $0.05$ for $100 \leq \rho \leq 200$, for larger values it slightly increases again to stabilize around $0.3$. Combining these considerations, the $\rho$ range of interest for detailed analysis has been selected in the interval $10\leq \rho \leq 100$.

\begin{figure}[h!]
\centering
\includegraphics[trim=2cm 0.1cm 2cm 0.1cm, clip=true,width=1\columnwidth]{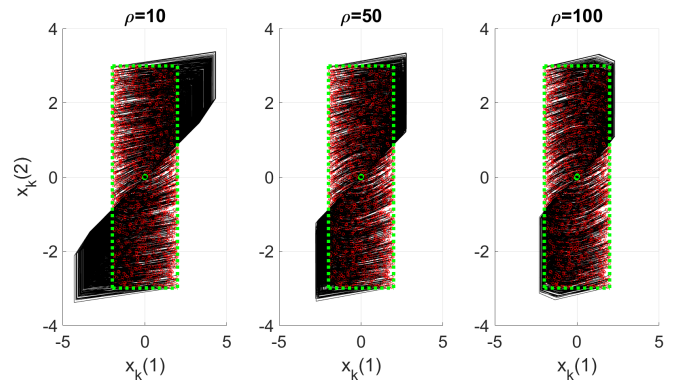}
\caption{State trajectories evolution for $\rho=[10,50,100]$ and $S_{\rho}=2614$ over the simulation horizon $M$.}
\label{f:traj_10_100}
\end{figure}

Fig. \ref{f:traj_10_100} provides an overview of the results for given values of $\rho$ in terms of state trajectories, each one starting from a random initial conditions $x_0$ (red circles). First, we can observe that for each of the considered values for $\rho$, there is at least one state trajectory that violates the constraints (green dotted square) but the violation seems to decrease when $\rho$ increases.

\begin{figure}[h!]
\centering
\includegraphics[trim=1.8cm 0.1cm 2cm 0.1cm, clip=true,width=1\columnwidth]{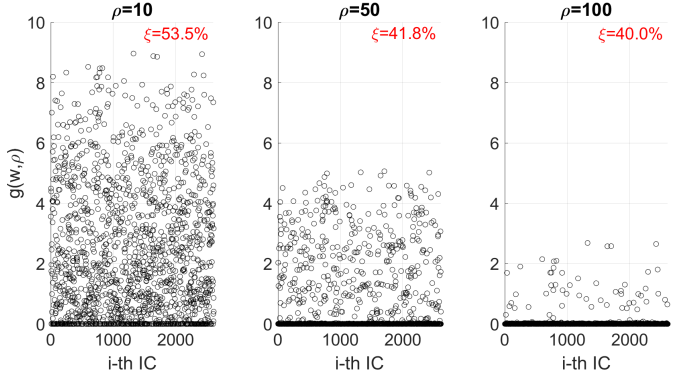}
\caption{Constraint violation entity over the simulation horizon $M$ for each of the $S_{\rho}$ initial conditions (IC) and for $\rho=[10,50,100]$. }
\label{fig:violation_10_100_17102019}
\end{figure}

Moreover, as supported also by Fig. \ref{fig:violation_10_100_17102019}, the number of trajectories that violate the constraint at least once significantly reduces while $\rho$ grows. Indeed, defined $\xi$ as the ratio between the number of trajectories violating the constraints $n_{viol}$ and the total number of simulations, i.e. $S_{\rho}$, we see that this percentage goes from above $53\%$ for $\rho=10$ to $\xi=40\%$ for $\rho=100$ but the attenuation of $g(w,\rho)$ results less effective with increasing $\rho$. In fact, from $\rho=10$ to $\rho=50$ we observe a $12\%$ reduction  whereas from $\rho=50$ to $\rho=100$ the reduction is smaller than $2\%$. 

\begin{figure}[h!]
\centering
\includegraphics[trim=0cm 0.0cm 0cm 0.0cm, clip=true,width=1\columnwidth]{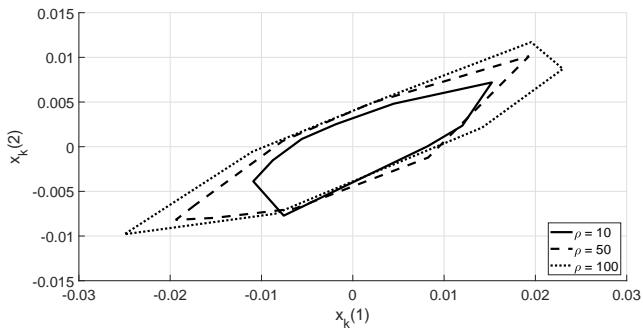}
\caption{Convex hull for last-step state trajectory $x_{M}$ for $\rho=[10,50,100]$.}
\label{fig:conv_hull_31102019}
\end{figure}

Focusing again on Fig. \ref{f:traj_10_100}, we can observe that in all cases the state trajectories converge to the origin (green circles). To understand how the terminal region changes with respect to the violation parameter, for each value of $\rho$ the convex hull of the terminal state $x_{M}$ has been obtained and represented in Fig. \ref{fig:conv_hull_31102019}. We can observe that increasing $\rho$, the terminal region seems to undergo an enlargement. The reason for this behaviour is that for smaller values of $\rho$ the controller focuses more on driving the initial conditions to the origin (at the expense of a larger probability of violation of the constraints). Notice that the controllers are less conservative for smaller values of $\rho$. Consequently, this leads also to smaller values of the nominal finite horizon cost $J(\textbf{z}_k,\textbf{v}_k)$ in \eqref{eq:J_final} and of the quadratic stage cost $L(x_k,u_k)$ in \eqref{eq:cost_L} along the closed-loop trajectory. For this particular example, the choice $\rho=100$ provides an adequate compromise between closed-loop performance (in terms of the quadratic stage cost $L(x_k,u_k)$) and constraint violation (expressed in terms of the violation level $\gamma(\rho)$ for $\epsilon_\rho=0.05$).

\section{Conclusions}
\label{sec:conclusion}
A stochastic model predictive controller able to account for the effects of additive stochastic disturbances is presented in this paper. No restrictive assumptions on the random nature of disturbances are required. We use a sampling method to bound offline the effect of disturbances in a probabilistic manner. A penalty based formulation, which avoids infeasibility of the optimization problem defining the model predictive controller, is proposed. The novel control scheme meets some given probabilistic closed-loop specifications. The required sample complexity has a logarithmic dependence with respect to the prediction horizon.  The efficacy of the proposed approach is demonstrated with a numerical example where the effects of the penalty factor on the controller are shown, providing a method to the users to select the best value of the penalty factor according to the application needs. 

\bibliography{ifacconf}             

\appendix
\section{Proof to \textit{Theorem 1}}
\label{app:app1}
Given $\ell \in \mathbb{N}_0^{N-1}$ and $j\in \mathbb{N}_1^{n_h}$, we denote  $E_{\ell,j}(\gamma)$ the probability of the event
$ \Pr_\setD\{ C_{K,j}e_{\ell|k}(\boldsymbol{\zeta}) > \gamma\}$. Property \ref{prop:max:gen} states that, with probability no smaller than
$$
1- \sum_{m=0}^{r_q-1} \conv{S_q}{m}\epsilon_q^m (1-\epsilon_q )^{S_q-m},
$$
we have 
\begin{eqnarray*}
E_{\ell,j}(q_{\ell,j})&=&  \Pr_\setD \{  C_{K,j}e_{\ell|k}(\boldsymbol{\zeta}) > q_{\ell,j}     \} \\
&=&   \Pr_\setD \{  C_{K,j}e_{\ell|k}(\boldsymbol{\zeta}) >  \maxgen{\{  C_{K,j}e_{\ell|k}(\boldsymbol{\zeta}^{(i)})\}_{i=1}^{S_q}}{r_q}  \} \leq \epsilon_q.
\end{eqnarray*}
That is,
$$ \Pr_{\setD^{S_q}}\{ E_{\ell,j}(q_{\ell,j})> \epsilon_q \}  \leq \sum_{m=0}^{r_q-1} \conv{S_q}{m}\epsilon_q^m (1-\epsilon_q )^{S_q-m}.$$

Consider now the probability $\delta_F$ that, after drawing $S_q$ i.i.d. samples $\{ \boldsymbol{\zeta}^{(i)} \}_{i=1}^{S_q}$, one or more of the obtained values for $q_{\ell,j}$ are not satisfying the constraint 
$$ E_{\ell,j}(q_{\ell,j}) = \Pr_\setD\{C_{K,j} e_{\ell|k}(\boldsymbol{\zeta}) > q_{\ell,j} \} \leq \epsilon_q.$$ 
We have 
 \begin{eqnarray*}
 \delta_F  &=&  \Pr_{\setD^{S_q}} \{ \epsilon_q < \max\limits_{\ell \in \N_0^{N-1}, j\in \N_1^{n_h}} E_{\ell,j}(q_{\ell,j})   \} \\ 
 & \leq & \sum_{\ell=0}^{N-1}\sum_{j=1}^{n_h} \Pr_{\setD^{S_q}}\{ \epsilon_q < E_{\ell,j}(q_{\ell,j})    \}\\ 
 & \leq & n_hN \sum_{m=0}^{r_q-1} \conv{S_q}{m}\epsilon_q^m (1-\epsilon_q )^{S_q-m} \leq \delta_q. 
 \end{eqnarray*}
 That is, $\delta_F\leq \delta_q$. This proves the first claim of the property. The second one follows directly from the second claim of Property \ref{prop:max:gen}. \QED
 
 \end{document}